# Singleshot Multispectral Imaging via a Chromatic Metalens Array

Romil Audhkhasi[1], Ningzhi Xie[1], Johannes E. Fröch[1] and Arka Majumdar[1,2,*]
[1]Department of Electrical and Computer Engineering, University of Washington, Seattle, Washington 98195, United States
[1]Department of Physics, University of Washington, Seattle, Washington 98195, United States
[*]Corresponding author: arka@uw.edu

**Abstract**
Real-time, single-shot multispectral imaging systems are crucial for environment monitoring and biomedical imaging. Most single-shot multispectral imagers rely on complex computational backends, which precludes real-time operations. In this work, we leverage the spectral selectivity afforded by engineered photonic materials to perform bulk of the multispectral data extraction in the optical domain, thereby circumventing the need for heavy backend computation. We use our imager to extract multispectral data for two real world objects at 8 predefined spectral channels in the 400 – 900 nm wavelength range. For both objects, an RGB image constructed using extracted multispectral data shows good agreement with an image taken using a phone camera, thereby validating our imaging approach. We believe that the proposed system can provide new avenues for the development of highly compact and low latency multispectral imaging technologies.

**Introduction**

The ability of multispectral imaging systems to retrieve the spatio – spectral data of a given scene makes them useful for a wide variety of applications ranging from defense and surveillance[1] to agriculture and food health monitoring.[2] A vast majority of the commercially available multispectral imagers utilize sensors that either scan across the spatial or the spectral domain of a scene to reconstruct its multispectral data cube.[3-5] However, the process of scanning causes such systems to have long data acquisition times.[6] A promising alternative is snapshot multispectral imaging that allows for data reconstruction from a single measurement, thus enabling faster data retrieval.[7]

In recent years, there has been a growing push to miniaturize multispectral imaging systems to allow for their integration into mobile platforms. Conventional multispectral imagers are often bulky and expensive, owing to the presence of multiple optical components and moving parts.[8-11] Photonic devices provide new avenues for the development of compact multispectral imagers due to their ability to achieve multiple optical functionalities within a compact form factor. The past few years have witnessed the development of several snapshot multispectral imaging systems based on both artificially engineered nanostructures[12-16] as well as commercially available random diffusers.[17, 18] While photonic snapshot multispectral imagers provide a significant reduction in form factor as compared to conventional scanning – based systems, their reliance on heavy computation prohibits real time operation and limits their practical utility.

In this work, we leverage the multi – functional nature of photonic devices to construct a multispectral imager that performs bulk of the spectral discrimination in the optical domain, thereby circumventing the need for heavy computation. Our imager is designed to retrieve the multispectral slices of an object at 8 wavelengths (referred to as spectral channels) in the 400 – 900 nm wavelength range. Figure 1 shows a schematic explaining the basic working principle of the imager. Light reflected by the object passes through a transmission filter with 8 narrow passbands in the 400 – 900 nm wavelength range. The filtered light illuminates an array of 8 metalenses, each of which is designed to focus light at one of the 8 spectral channels. The metalens array produces spatially separated images of the multispectral slices of the object at its focal plane that are captured by a monochrome camera sensor.

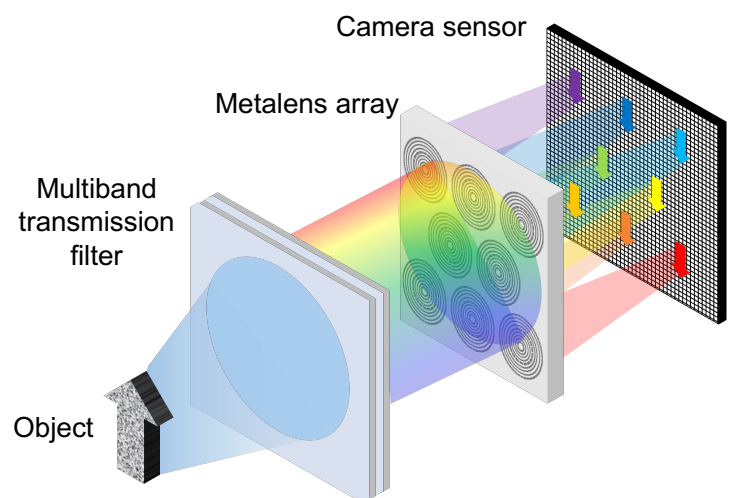

**Figure 1:** Schematic showing the working principle of our multispectral imager.

While the inherent chromaticity of metalenses is considered to be undesirable for most imaging applications, here we leverage it to achieve a spatial separation of the scene's multispectral data. This enables data retrieval without the need for significant computational post – processing. In the next section, we provide a detailed description of the structure and functioning of the components of our multispectral imager, namely the multiband transmission filter and the metalens array. Subsequently, we use our imager to retrieve multispectral data for two real – world objects in the visible and near – infrared wavelength range. We believe that our imaging approach provides new avenues for the development of highly compact snapshot multispectral imagers for real – time operation.

**Components of the multispectral imager**

As shown in Fig. 1, the imager comprises of a multiband transmission filter and an array of strongly chromatic metalenses. The purpose of the transmission filter is to isolate spatial information from a given scene corresponding to 8 wavelengths in the 400 – 900 nm range. To achieve this, we design a filter whose transmission spectrum contains multiple narrow resonances in the wavelength range of operation. The filter consists of a 50 nm thick layer of $SiO_2$ sandwiched between two chirped $Si_3N_4$ – $SiO_2$ distributed Bragg reflectors (DBRs) with 21 layers each (Fig. 2(a)). Given the low refractive index contrast between $Si_3N_4$ and $SiO_2$, a conventional, periodic DBR formed from these materials is not expected to have a stopband that extends over the entire 400 – 900 nm wavelength range. In order to achieve broadband reflection, we use chirped DBRs in which the layer thickness increases linearly from top to bottom. The thickness of the $i^{th}$ layer is chosen to be $\lambda_i / 4n_i$, where $\lambda_i$ is one of 21 equally spaced wavelengths between 400 and 900 nm and $n_i$ is the layer's refractive index. While designing the filter, we neglect material loss and dispersion, and consider the refractive indices of $Si_3N_4$ and $SiO_2$ as 2.04 and 1.45, respectively.

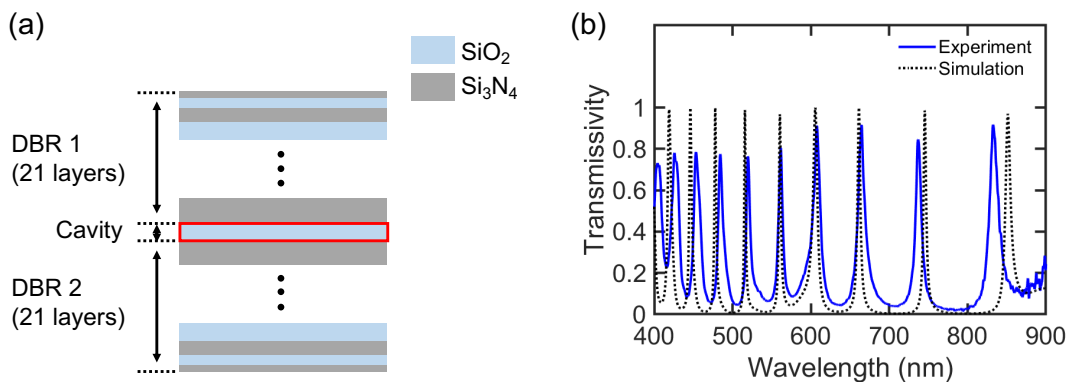

**Figure 2:** (a) Schematic of the chirped – DBR based multiband transmission filter. (b) Simulated and experimentally measured transmission spectra of the filter.

We fabricate the 43 – layer multiband transmission filter by alternatingly depositing $Si_3N_4$ and $SiO_2$ on a fused silica chip using plasma enhanced chemical vapor deposition (SPTS Technologies Ltd., Delta LPX). Figure 2(b) shows the measured transmission spectrum of the multiband filter (solid blue curve). We observe that the filter exhibits 10 narrow transmission resonances in the 400 – 900 nm wavelength range. We use the peak locations of the 8 resonances located above 450 nm to define the channels at which multispectral data is collected in this study. For comparison, we also show the filter's transmission spectrum calculated using the transfer matrix formalism (dotted black curve). While the two spectra are in good agreement, the resonances in the measured spectrum are slightly blue shifted and approximately twice as broad as those in the calculated spectrum. This can be attributed to the presence of loss in the actual materials and slight deviations in their refractive indices from those assumed during calculations.

The second component of our multispectral imager is an array of 8 strongly chromatic metalenses (Fig. 3(a)). The array is designed such that scene information corresponding to the 8 spectral channels of interest is imaged at spatially separated locations on its focal plane. To enable this, we design each metalens of the array as a 4 mm diameter hyperboloid that focuses incident light at one of the 8 spectral channels at a distance of 10 mm from it (f/# = 2.5). We note that while the multiband filter transmits scene information corresponding to all of its 8 passbands, the metalenses only image spatial content at their respective operating wavelengths on to the focal plane. Therefore, unlike other imaging applications, our multispectral imager leverages the inherent chromaticity of conventional hyperboloid metalenses to achieve additional wavelength selectivity over the multiband transmission filter.

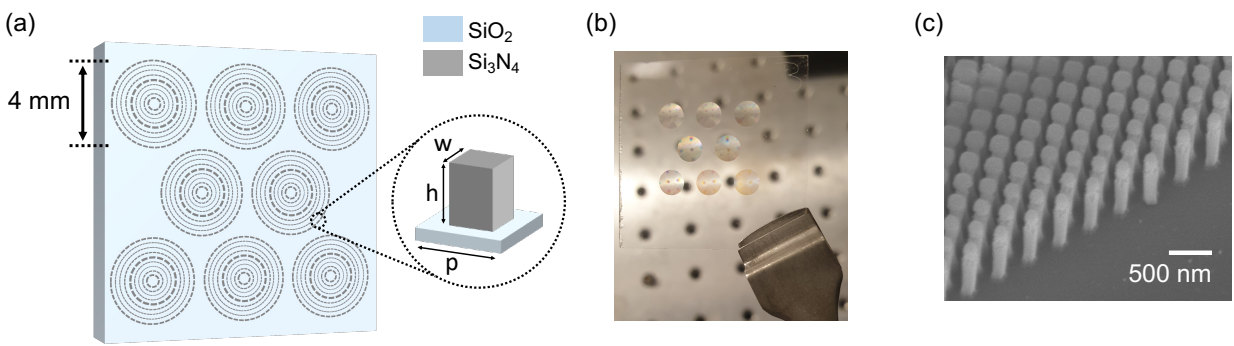

**Figure 3:** (a) Schematic of the metalens array with a single constituent nanopillar shown in the inset. (b) Image of the fabricated metalens array taken with a phone camera. (c) Scanning electron microscope image of one of the metalenses.

Each metalens consists of square $Si_3N_4$ nanopillars of height $h$ = 1 µm placed on a $SiO_2$ substrate with a center-to-center separation of $p$ = 300 nm (inset of Fig. 3(a)). The widths of the nanopillars in a given metalens are chosen to achieve a hyperbolic phase profile at the corresponding operating wavelength. The metalens array is fabricated using standard electron beam lithography (see 'Methods' for details of the metalens design and fabrication procedures). Figure 3(b) shows an image of the fabricated metalens array taken with a phone camera while 3(c) shows a scanning electron microscope image of one of the metalenses.

In the next section, we use our imager to retrieve multispectral information for two real – world objects in the 400 – 900 nm wavelength range.

**Results and discussion**

As the first test object for our system, we consider a standard color chart target (Fig. 4(a)). The chart comprises of 24 reflective colored squares and is commonly used for calibrating imaging systems in photography. Section S1 of the Supporting Information provides details of our experimental setup for retrieving the multispectral slices of a given object. We note that our imaging approach does not require the presence of any optics other than the filter and lens array and hence provides a simplified route to multispectral data extraction. Figure 4(b) shows an image of the object recorded by using our multispectral imager with a monochrome camera. The image shows 8 spatially separated copies of the object with varying intensities (labeled 1 through 8). Each copy corresponds to a multispectral slice of the object with additional artefacts arising due to the metalens that was used to image it.

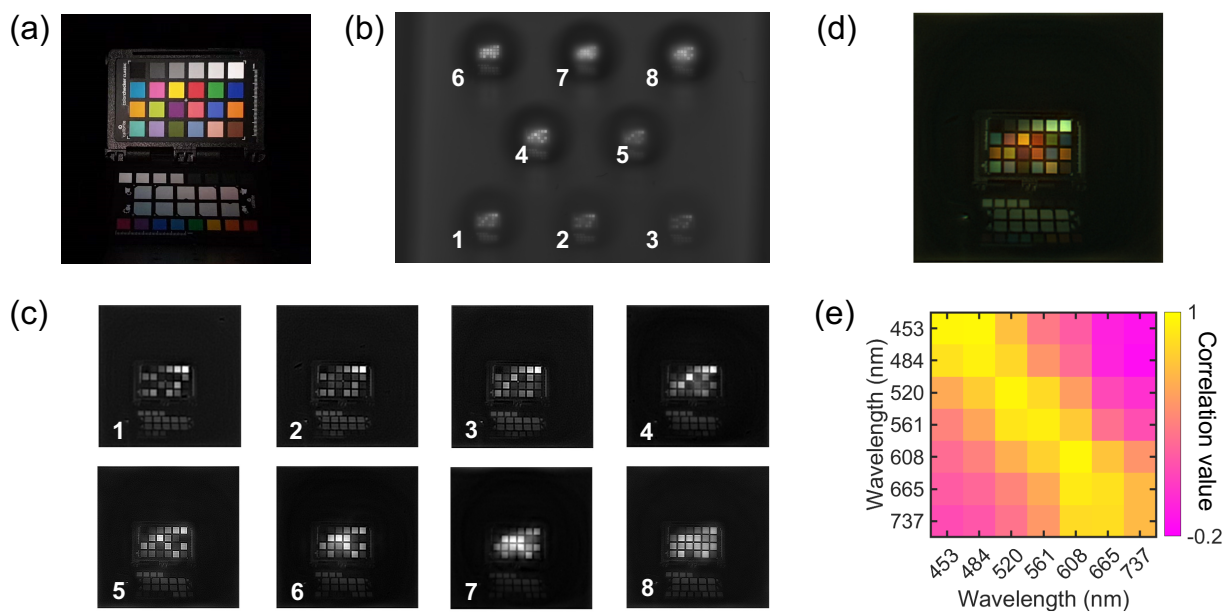

**Figure 4:** (a) Image of a color chart target object taken with a phone camera. (b) Image of the object produced by our multispectral imager. (c) Processed multispectral slices of the object. (d) RGB image of the object obtained by combining its reconstructed RGB channels. (e) Correlation matrix between the spatial reflectivity maps of the first 7 multispectral slices with those obtained from catalogued spectra. The *x* and *y* axes indicate the wavelengths corresponding to the 7 slices.

To recover more accurate multispectral information of the given object, we isolate the individual slices from the camera image and post-process them. Section S2 of the supporting Information provides details of the computational backend used to improve the quality of the captured slices using Wiener deconvolution and haze reduction. The processed multispectral slices of the color chart object are presented in Fig. 4(c). To check the validity of our multispectral imaging approach, we use the 8 slices of the object with a color mapping function[19] to retrieve its RGB channels. The image of the object obtained by combining the reconstructed RGB channels (Fig. 4(d)) matches qualitatively with the image taken by a phone camera (Fig. 4(a)). We further quantify the correctness of our approach by calculating the correlation between the spatial reflectivity maps of the 8 slices with those obtained from catalogued spectra of the color chart (see section S3 of the Supporting Information for details). Figure 4(d) shows the resulting 7x7 correlation matrix for 7 of the 8 slices due to non-availability of the catalogued spectra above a wavelength of 780 nm. The band diagonal nature of the matrix validates the multispectral data obtained using our system.

For the second demonstration of our multispectral imaging approach, we use a scene consisting of flowers and leaves (Fig. 5(a)). Figure 5(b) presents the processed multispectral slices of the scene obtained using the method as used in Fig. 4. We observe that the spectral content of the scene is consistent with the colors of the various objects contained in it. For instance, the orange flower appears brighter at longer wavelengths (slices 3 through 8) due to the presence of the colors yellow and red while the pink flower is also visible at shorter wavelengths due to its constituent color blue. We note that the green leaf shows up faintly in all the multispectral slices as it is not properly illuminated in the original scene. Using a color mapping function,[19] we reconstruct the R, G and B channels of the object and combine those to create an RGB image (Fig. 5(c)). This image shows good qualitative agreement with the image of the object captured using a phone camera (Fig. 5(a)), thus providing further evidence for the validity of our multispectral imaging approach.

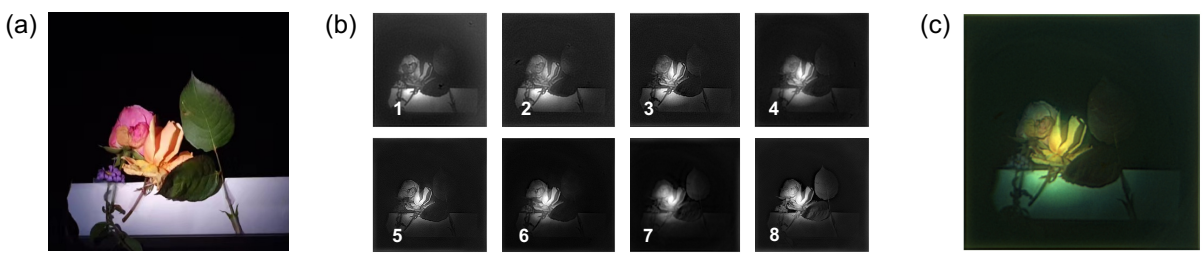

**Figure 5:** (a) Image of flowers and leaves taken with a phone camera. (b) Processed multispectral slices of the object. (c) RGB image of the object obtained by combining its reconstructed RGB channels.

## **Conclusion**

In this work, we proposed an 8 – channel multispectral imager operating in the 400 – 900 nm wavelength range. The imager first utilizes a multi – resonant transmission filter to isolate the spatial content of a given scene contained within 8 narrow passbands centered at the wavelength channels of interest. Then, an array of 8 metalenses designed to individually focus light at one of the 8 wavelength channels is used to form spatially separated images of the scene's spectral content on a monochrome camera sensor. Such a spatial separation of the spectral information allows our system to retrieve multispectral data with minimal computational post – processing. We tested the validity of our scheme by retrieving multispectral slices for two different real – world objects and using those to reconstruct their respective R, G and B channels. The colored images of the objects obtained by combining their respective three-channel data were in good agreement with their images taken with a standard phone camera.

We note that the proposed multispectral imaging system suffers from two major limitations. First, for a general $N$ – channel system, each metalens of the array receives at most $1/N^{th}$ of the power transmitted by the multiband filter. Of the total received power, a given metalens only focuses power corresponding to its wavelength of operation on to the camera sensor while transmitting the rest. This significantly reduces the signal-to-noise ratio of the captured multispectral data. Furthermore, for a given camera sensor, the spatial resolution of the recovered multispectral data is lower than that of the original scene by an amount dependent on the number of wavelength channels being retrieved.

Despite the above limitations, the proposed system illustrates a convenient method for real – time multispectral data acquisition. Future work may explore alternatives to the metalens array for power efficient spatial separation of spectral data. For example, instead of creating an array of multiple monochromatic metalenses, one may design a single, large area meta – optic that focuses incoming light at different spatial locations based on its wavelength. Such a device would allow different spectral components of a given scene to be imaged at different spatial locations on the camera sensor with minimal crosstalk thus enabling computation – free data retrieval.

## Methods
**Design of the metalens array**
Each metalens of the array is designed to impart a hyperbolic phase profile to a normally incident plane wave at its wavelength of operation. The phase delay due to a point on the metalens at a distance $r$ from its center is given by the following equation:

$$\varphi(r) = -\frac{2\pi}{\lambda}\left(\sqrt{f^2 + r^2} - f\right)$$

Here $\lambda$ is the operating wavelength of the lens (corresponding to one of the 8 spectral channels) and $f$ is its focal length fixed at 10 mm.

We begin the design process by constructing a library of $Si_3N_4$ nanopillars with a height of 1 μm, period of 300 nm, and varying widths. The library consists of the wavelength – dependent transmission amplitudes and phases of the nanopillars calculated using Rigorous Coupled Wave Analysis (RCWA) simulations. Next, we bin the circularly symmetric phase profile of each metalens into 8 levels. For each level, we select a meta-atom from the library that imparts the required phase at the lens's operating wavelength. The lenses are formed by placing the chosen nanopillars at their respective radii.

**Fabrication of the metalens array**
The metalens is fabricated using an electron beam lithography (EBL) process. A 1 μm thick $Si_3N_4$ layer is deposited on a 0.5mm thick fused silica chip using plasma enhanced chemical vapor deposition (SPTS Technologies Ltd., Delta LPX). An e-beam resist (ZEP-520A) is then spin-coated onto the chip at 5000 rpm and patterned by EBL (JEOL Ltd., JBX-6300FS). Subsequently, a 100 nm thick $Al_2O_3$ hard mask is created by electron beam assisted evaporation (CHA Industries, SEC-600) on the patterned resist followed by resist lift-off in N-methyl-2-pyrrolidone (NMP) solution at 90 degree Celsius for 8 hours. Subsequently, the $Si_3N_4$ layer is etched via a fluorine based reactive ion etch process (Oxford, PlasmaLab 100, ICP-180).

**Acknowledgements**
This work was supported by the DARPA Coded Visibility STTR program.

**Supporting Information**
Experimental setup for multispectral data extraction, computational post-processing of multispectral data and correlation matrix calculations.

**References**
(1) Shimoni, M.; Halelterman, R.; Perneel, C. Hyperspectral Imaging for Military and Security Applications: Combining Myriad Processing and Sensing Techniques. *IEEE Geoscience and Remote Sensing* **2019**, *7* (2), 101 - 117.
(2) Liu, Y.; Pu, H.; Sun, D.-W. Hyperspectral imaging technique for evaluating food quality and safety during various processes: A review of recent application. *Trends in Food Science & Technology* **2017**, *69*, 25 - 35.
(3) Green, R. O.; Eastwood, M. L.; Sarture, C. M.; Chrien, T. G.; Aronsson, M.; Chippendale, B. J.; Faust, J. A.; Pavri, B. E.; Chovit, C. J.; Solis, M.; et al. Imaging Spectroscopy and the Airborne Visible/Infrared Imaging Spectrometer (AVIRIS). *Remote Sensing of Environment* **1998**, *65* (3), 227 - 248.
(4) Gat, N. *Imaging spectroscopy using tunable filters: a review*; SPIE, 2000.
(5) Zhang, C.; Rosenberger, M.; Breitbarth, A.; Notni, G. A novel 3D multispectral vision system based on filter wheel cameras. *IEEE International Conference on Imaging Systems and Techniques (IST)* **2016**, 267 - 272. DOI: 10.1109/IST.2016.7738235.
(6) Cao, X.; Yue, T.; Lin, X.; Lin, S.; Yuan, X.; Dai, Q.; Carin, L.; Brady, D. J. Computational Snapshot Multispectral Cameras: Toward dynamic capture of the spectral world. *IEEE Signal Processing Magazine* **2016**, *33* (5), 95 - 106.
(7) Gao, L.; Wang, L. V. A review of snapshot multidimensional optical imaging: Measuring photon tags in parallel. *Physics Reports* **2016**, *616*, 1 - 37.
(8) Kar, O. F.; Oktem, F. S. Compressive spectral imaging with diffractive lenses. *Optics Letters* **2019**, *44* (18), 4582 - 4585.
(9) Sullenberger, R. M.; Milstein, A. B.; Rachlin, Y.; Kaushik, S.; Wynn, C. M. Computational reconfigurable imaging spectrometer. *Optics Express* **2017**, *25* (25), 31960 - 31969.
(10) Wagadarikar, A.; John, R.; Willett, R.; Brady, D. Single disperser design for coded aperture snapshot spectral imaging. *Applied Optics* **2008**, *47* (10), 844 - 851.
(11) Gehm, M. E.; Brady, J. D. J.; Willett, R. M.; Schulz, T. J. Single-shot compressive spectral imaging with a dual-disperser architecture. *Optics Express* **2007**, *15* (21), 14013 - 14027.
(12) French, R.; Gigan, S.; Muskens, O. L. Speckle-based hyperspectral imaging combining multiple scattering and compressive sensing in nanowire mats. *Optica Letters* **2017**, *42*, 1820 - 1823.

(13) Jeon, D. S.; Baek, S.-H.; Yi, S.; Fu, Q.; Dun, X.; Heidrich, W.; Kim, M. H. Compact snapshot hyperspectral imaging with diffracted rotation. *ACM Trans. Graph.* **2019**, *38* (4), Article 117. DOI: 10.1145/3306346.3322946.
(14) Baek, S.-H.; Ikoma, H.; Jeon, D. S.; Yuqi, L.; Heidrich, W.; Wetzstein, G. Single-shot Hyperspectral-Depth Imaging with Learned Diffractive Optics. *IEEE/CVF International Conference on Computer Vision* **2021**, 2631 - 2640. DOI: 10.1109/ICCV48922.2021.00265.
(15) August, I.; Oiknine, Y.; AbuLeil, M.; Abdulhalim, I.; Stern, A. Miniature Compressive Ultra-spectral Imaging System Utilizing a Single Liquid Crystal Phase Retarder. *Scientific Reports* **2016**, *6*, 23524.
(16) Arguello, H.; Pinilla, S.; Peng, Y.; Ikoma, H.; Bacca, J.; Wetzstein, G. Shift-variant color-coded diffractive spectral imaging system. *Optica* **2021**, *8* (11), 1424 - 1434.
(17) Monakhova, K.; Yanny, K.; Aggarwal, N.; Waller, L. Spectral DiffuserCam: lensless snapshot hyperspectral imaging with a spectral filter array. *Optica* **2020**, *7*, 1298 - 1307.
(18) Sahoo, S. K.; Tang, D.; Dang, C. Single-shot multispectral imaging with a monochromatic camera. *Optica* **2017**, *4*, 1209 - 1213.
(19) *Convert wavelength of light in representative color*; GitHub, 2020. https://github.com/razanskylab/wavelength2rgb/releases/tag/1.1 (accessed.

# Supporting information for "Singleshot Multispectral Imaging via a Chromatic Metalens Array"

Romil Audhkhasi[1], Ningzhi Xie[1], Johannes E. Fröch[1] and Arka Majumdar[1,2,*]

[1]Department of Electrical and Computer Engineering, University of Washington, Seattle, Washington 98195, United States

[1]Department of Physics, University of Washington, Seattle, Washington 98195, United States

[*]Corresponding author: arka@uw.edu

Number of pages: 3
Number of figures: 3

## S1. Experimental setup for multispectral data extraction

Figure S1 shows an image of the experimental setup used for retrieving the multispectral slices of a given object. Light from a broadband source reflected by the object passes through the metalens array and the multiband transmission filter. For convenience, the metalens array is taped on to the transmission filter which in turn is mounted on a translation stage. The image formed by the metalens array on its focal plane is captured by using a monochrome camera sensor. We note that our imaging approach does not require the presence of any optics other than the filter and lens array and hence provides a simplified route to multispectral data extraction.

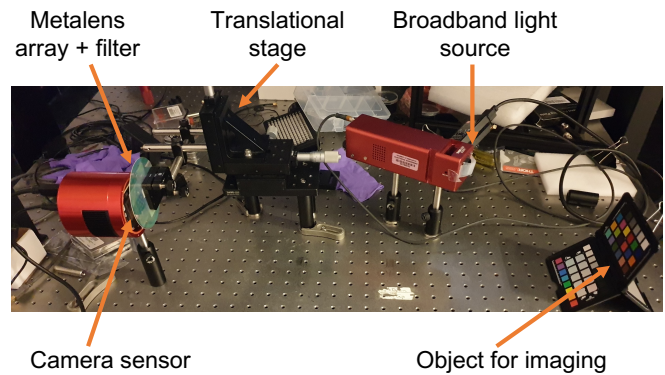

**Figure S1:** Experimental setup for retrieving the multispectral slices of a given object.

## S2. Computational post – processing of multispectral data

Prior to post – processing, an image of the illuminating broadband light source is recorded using the multispectral imager. This image comprises of 8 spatially separated focal spots corresponding to the point spread functions (PSFs) of the metalenses at their respective operating wavelengths. Figure S2 presents a schematic showing the computational post – processing of slice no. 5. First, the captured image of the slice undergoes Wiener deconvolution with the PSF of its corresponding metalens using MATLAB's inbuilt deconvwnr function. To remove haze in the deconvolved slice, we use MATLAB's inbuilt imreducehaze function which implements the approximate dark channel prior algorithm for haze reduction.[1] The processed multispectral slices of the color chart object are presented in Fig. 4(c) of the main manuscript.

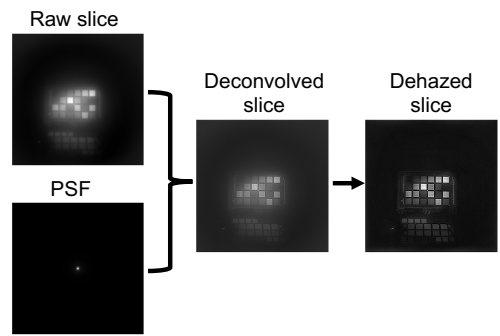

**Figure S2:** Post – processing of a single multispectral slice.

## S3. Correlation matrix calculations

To obtain the correlation matrix, we first determine the reflectivities of the 24 squares of the color chart object from its processed multispectral slices. Figure S3 shows this process for slice no. 2. We assume a rectangular region around the center of each square and calculate the mean value of the pixels inside it. This gives the reflectivity of each of the 24 squares at the wavelength corresponding to slice 2. Repeating this process for all the slices gives the reflectivity of each square (reflectivity map of the object) at each of the 8 wavelengths. The ground truth reflectivity map is obtained by using catalogued reflectivity values of the 24 squares. The ($i$, $j$)th element of the correlation matrix corresponds to the Pearson correlation coefficient between the 24-element reflectivity map at the $i$th wavelength channel of the ground truth and $j$th wavelength channel of the calculated multispectral data.

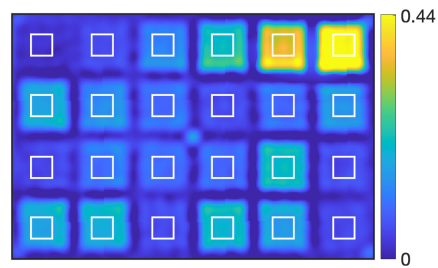

**Figure S3:** Processed slice no. 2 showing rectangular regions that are used to calculate the reflectivity of each square.

## References

(1) He, K.; Sun, J.; Tang, X. Single image haze removal using dark channel prior. *IEEE Conference on Computer Vision and Pattern Recognition* **2009**, 1956 - 1963. DOI: 10.1109/CVPR.2009.5206515.